\documentclass[useAMS,usenatbib]{mnras}

\usepackage{url,times,hyperref,graphicx,amsmath,amsfonts,amssymb,aas_macros,color,epsfig,epstopdf}

\newcommand{\mhalo}{M$_{\rm halo}$}
\newcommand{\mstar}{M$^{\star}$}
\newcommand{\mHI}{M$_{\rm HI}$}
\newcommand{\lcdm}{$\Lambda$CDM}
\newcommand{\msun}{M$_\odot$}

\newcommand{\hMpc}{{\ifmmode{h^{-1}{\rm Mpc}}\else{$h^{-1}$Mpc}\fi}}
\newcommand{\hkpc}{{\ifmmode{h^{-1}{\rm kpc}}\else{$h^{-1}$kpc}\fi}}
\newcommand{\hMsun}{{\ifmmode{h^{-1}{\rm {M_{\odot}}}}\else{$h^{-1}{\rm{M_{\odot}}}$}\fi}}
\newcommand{\ltsima}{$\; \buildrel < \over \sim \;$}
\newcommand{\gtsima}{$\; \buildrel > \over \sim \;$}
\newcommand{\lsim}{\lower.5ex\hbox{\ltsima}}
\newcommand{\gsim}{\lower.5ex\hbox{\gtsima}}
\def\lcdm{$\Lambda$CDM}

\def\lesssim{\mathrel{\hbox{\rlap{\hbox{\lower4pt\hbox{$\sim$}}}\hbox{$<$}}}}
\def\gtrsim{\mathrel{\hbox{\rlap{\hbox{\lower4pt\hbox{$\sim$}}}\hbox{$>$}}}}

\newcommand{\Sec}[1]{Section~\ref{#1}}

\newcommand{\Fig}[1]{Fig.~\ref{#1}}
\newcommand{\beq}{\begin{equation}}
\newcommand{\eeq}{\end{equation}}
\def\beqa{\begin{eqnarray}}
\def\eeqa{\end{eqnarray}}
\def\hMpc{$h^{-1}\,{\rm Mpc}$}
\def\hkpc{$h^{-1}\,{\rm kpc}$}
\def\lcdm{\ensuremath{\Lambda}CDM}

\def\head{
 \vbox to 0pt{\vss
                   \hbox to 0pt{\hskip 440pt\rm LA-UR-10-07069\hss}
                  \vskip 25pt}}

\title[UDGs are expanded dwarfs]
{NIHAO XI: Formation of Ultra-Diffuse Galaxies by outflows} 
\author[Di Cintio]
       {Arianna Di Cintio$^{1}$\thanks{E-mail: arianna.dicintio@dark-cosmology.dk}, Chris B. Brook$^2$, Aaron A. Dutton$^{3}$, Andrea V. Macci\`{o}$^{3,4}$, \newauthor  Aura Obreja$^{3}$ \& Avishai Dekel$^5$\\
$^{1}$Dark-Carlsberg Fellow, Dark Cosmology Centre, NBI, University of Copenhagen, Juliane Maries Vej 30, DK-2100 Copenhagen, Denmark\\
$^{2}$Ramon y Cajal Fellow, Universidad de La Laguna and Instituto de Astrof\'isica de Canarias, E-38206, La Laguna, Tenerife, Spain\\
$^{3}$New York University Abu Dhabi, PO Box 129188, Abu Dhabi, United Arab Emirates\\
$^{4}$Max Planck Institute f\"{u}r Astronomie, K\"{o}nigstuhl 17, 69117 Heidelberg, Germany\\
$^5$Center for Astrophysics and Planetary Science, Racah Institute of Physics, The Hebrew University, Jerusalem 91904, Israel}

\begin{document}

\date{Accepted xxxx. Received xxxx; in original form xxxx}

\pagerange{\pageref{firstpage}--\pageref{lastpage}} \pubyear{2010}

\maketitle

\label{firstpage}

\begin{abstract}
We address the origin of  Ultra-Diffuse Galaxies (UDGs), which have stellar masses typical of dwarf galaxies but effective radii  of Milky Way-sized objects. Their formation mechanism, and  
whether they  are failed $\rm L_{\star}$ galaxies or  diffuse dwarfs, are challenging issues.
Using zoom-in cosmological simulations from the NIHAO project, we show that UDG analogues form naturally in dwarf-sized haloes due to episodes of gas outflows associated with star formation. The simulated UDGs live in isolated haloes of masses $10^{10-11}$\msun, have stellar masses of $10^{7-8.5}$\msun, effective radii larger than 1\,kpc and dark matter cores. They show a broad range of colors, an average S\'ersic index of 0.83, a typical distribution of halo spin and concentration, and a non-negligible HI gas mass of $10^{7-9}$\msun, which correlates with the extent of the galaxy.
Gas availability  is crucial to the internal processes that form UDGs: feedback driven gas outflows, and  subsequent dark matter and stellar expansion, are the key to reproduce faint, yet unusually extended, galaxies.
This scenario implies that UDGs represent a dwarf population of low surface brightness galaxies and should exist in the field. The largest  isolated UDGs should contain more HI gas than less extended dwarfs of similar \mstar.
\end{abstract}

\noindent
\begin{keywords}
 galaxies: dwarf - evolution - formation - haloes 
 \end{keywords}

\vspace{-1cm}\section{Introduction} \label{sec:introduction}
Deep imaging of nearby clusters have revealed a sizeable  population of faint, $M_{\rm R}$$\gtrsim$$-16.5$, low surface brightness, $\mu_e$=24-28 mag/arcsec$^2$, and unusually large, $0.8$$<$$r_e/\rm kpc$$<$$5$ galaxies, named Ultra-Diffuse Galaxies or UDGs \citep{vandokkum15}. While their stellar masses are typical of dwarfs, $10^7$$<$\mstar/\msun$<$$10^{8.7}$, their effective radii are compatible with $\rm L_{\star}$ objects, raising doubts about the  nature of such galaxies.
Almost 1000 UDGs have been identified in the Coma cluster using the Dragonfly array \citep{vandokkum15,vandokkum15b} and the Subaru telescope \citep{koda15,yagi16}: they represent a passively evolving population, lying on the red sequence in the color-magnitude diagram, as opposed to classical low surface brightness (LSB)  galaxies which are bluer and brighter \citep[e.g.][]{mcgaugh95,impey97,schombert11}. 
Red UDGs have since been found  in the Virgo cluster \citep{mihos15}, with even lower SB than the  Virgo LSBs discovered 30 years ago by \citet{sandage84}. Confirming the existence of an abundant population of such objects, UDGs have also been observed in Fornax and other clusters \citep{munoz15,vanderburg16}.
However, UDGs may not necessarily be red and associated  with clusters: \citet{roman16} studied a  region around the Abell 168 cluster, showing that about half of its UDGs are found outside the main cluster overdensity, with their properties changing towards the cluster centre, suggestive of environmental effects; they also observed  bluer UDGs than in Coma. Corroborating the idea that UDGs can  form in isolation, \citet{delgado16} found a UDG in the outskirts of the Pisces-Perseus supercluster, DGSAT I, that shows a blue over-density compatible with recent star formation.\\
\indent While some authors envision a scenario in which UDGs are failed $\rm L_{\star}$ galaxies that lost their gas after forming the first stars \citep{vandokkum15,vandokkum15b}, some others argue that they are genuine dwarf galaxies possibly living in high-spin haloes \citep{amorisco16}. Supporting the first claim are simulations by \citet{yozin15}, indicating that UDGs may be underdeveloped galaxies whose early accretion onto a cluster quenched further growth.
Favouring a massive halo for UDGs is  the  inferred virial mass of $\sim$8$\times$10$^{11}$\msun\ for DF 44\footnote{The \citet{maccio08} c-M relation was used, giving  up to 30$\%$ lower concentration than the Planck one used here, allowing a fit  into a larger halo.}, one of the brightest  Coma UDGs  \citep{vandokkum16}. This result is  in contrast with the  derived mass of two other  UDGs:  using the abundance of their globular clusters, \citet{beasley16}  inferred a \mhalo\ of (8$\pm$4)$\times$10$^{10}$\msun\  for VCC 1287, while \citet{Peng16}  and \citet{beasley16b} determined a total mass of  ($\sim$9$\pm$2)$\times$10$^{10}$\msun\ for DF17, favouring the idea that UDGs are dwarfs. Further, \citet{roman16}  showed that the spatial distribution of UDGs in  Abell 168  is compatible with the one of standard dwarfs.

A key question is whether UDGs can arise within a \lcdm\ universe.
An appealing possibility is that the formation of UDGs is not connected to the cluster environment, but rather to internal processes, such that  UDGs already have a spatially extended stellar component when infalling into a cluster. Previous simulation works extensively showed that feedback driven gas outflows are able to cause  expansion not only of the central DM distribution in galaxies \citep[e.g.][]{governato10,DiCintio2014a,DiCintio2014b,onorbe15,tollet15,chan15,read16}, but also of the stellar one \citep[e.g.][]{teyssier13,elbadry16,dutton16}.
The formation  of DM density cores  is related to rapid oscillation of the central potential driven by gas outflows following bursty star formation \citep[e.g.][]{read05,mash08,pontzen12} and  has been  applied to observations to reconcile the \textit{cusp-core} discrepancy \citep{katz16}. 
Interestingly, the mass range where we expect maximum efficiency in core formation overlaps with that of UDGs, i.e. galaxies with \mstar$\sim$$10^{7-9}$\msun\ should form large DM  and  stellar cores, while at higher and lower masses energy from stellar feedback alone becomes less efficient at creating  cores \citep{DiCintio2014a,DiCintio2014b,dutton16}. 
It is thus natural to explore whether feedback driven expansion is a viable mechanism for the formation of UDGs. 
In this \textit{Letter} we show that  isolated UDGs, with a spatially extended stellar distribution,  form naturally in dwarf-sized haloes  by  gas outflows. In \Sec{sec:simulations_NIHAO} we introduce the hydrodynamical cosmological simulations used, in \Sec{sec:results} we  investigate the formation scenario of UDGs and show results focusing on global properties and gas content of UDGs, and in \Sec{sec:discussion} we conclude by highlighting some observational predictions of our model.
\vspace{-.3in}\section{Simulations} \label{sec:simulations_NIHAO}
The  simulated galaxies are taken from the Numerical Investigation of a Hundred Astrophysical Objects (NIHAO) project \citep{wang15},  evolved using  the SPH code Gasoline  \citep{wadsley04,keller14}. The code includes a subgrid model for turbulent mixing of metals and energy \citep{wadsley08}, ultraviolet  heating, ionization and metal cooling \citep{shen10}.
Star formation and feedback follows the model used in the MaGICC simulations \citep{stinson13}, that for the first time reproduced several galaxy scaling relations over a wide mass range \citep{brook12b},
adopting a threshold for star formation of $n_{\rm th}$$>$$10.3 \rm cm^{-3}$. 

Stars feed energy back into the ISM via blast-wave supernova  feedback \citep{stinson06} and early stellar feedback from massive stars. Particle masses and force softenings  are chosen to resolve the mass profile to below 1$\%$ of the virial radius at all masses, ensuring that galaxy half-light radii are  well resolved.
The NIHAO galaxies cover a broad mass range, from dwarfs to Milky Way mass, and represent an unbiased sampling of merger histories, concentrations and spin parameters. The  galaxies are all centrals and isolated, and lie  on  abundance matching predictions, having the expected \mstar\ for each \mhalo.
The NIHAO project satisfactorily reproduces  realistic galaxies in terms of their \mstar, SFH, metals and DM distribution \citep[e.g.][]{tollet15,obreja16}.  

The  haloes are identified using the AHF 
halo finder \citep{Knollmann09} and partially analysed with the \textit{pynbody} package \citep{Pontzen13}.  

  \begin{figure}
\hspace{.7cm}\includegraphics[width=2.8in,height=4.9in]{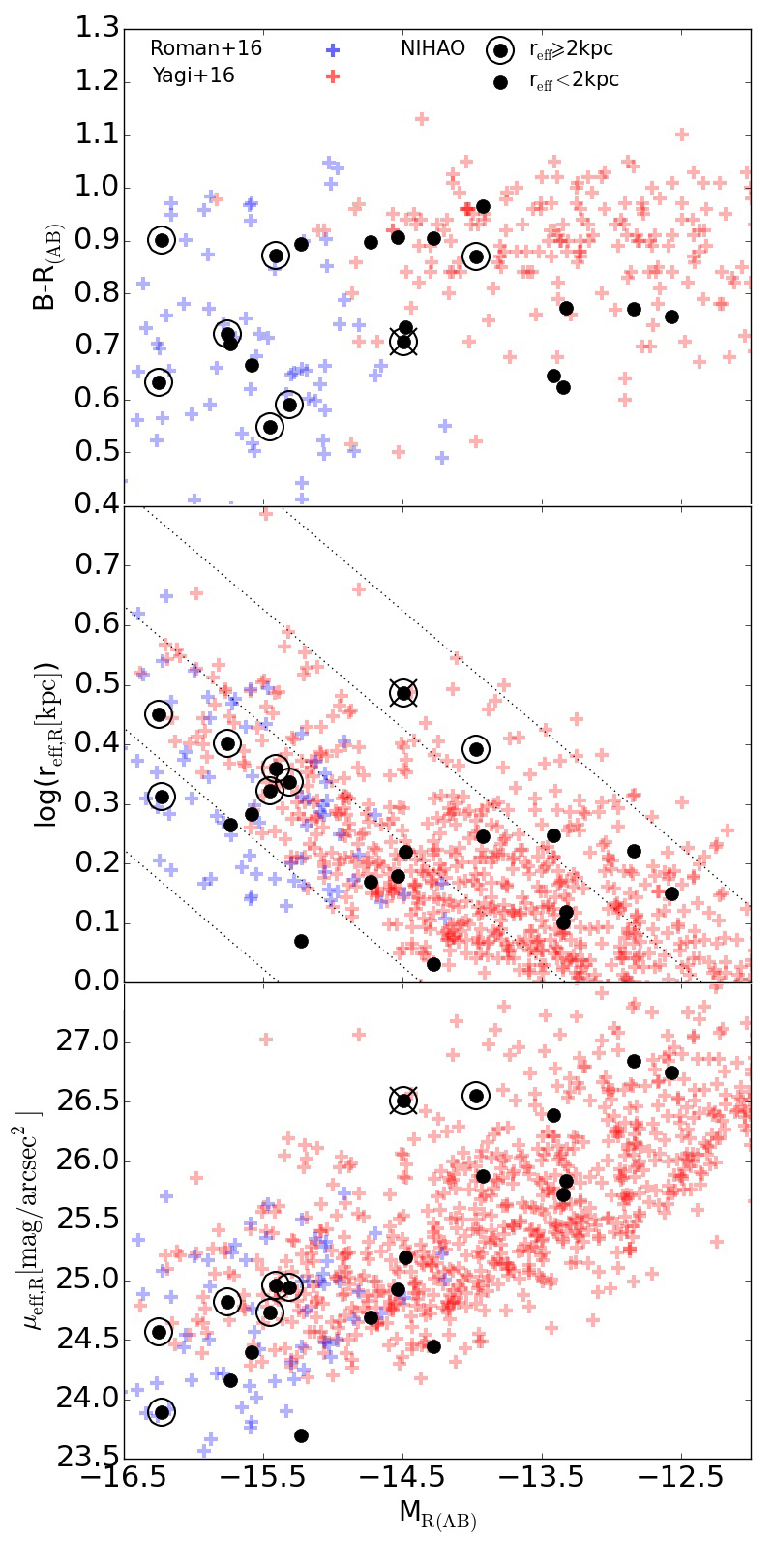}
  \caption{From top to bottom, we show B-R, $r_e$ and $\mu_e$ as a function of R-band absolute magnitude of Coma cluster UDGs from \citet{yagi16} (red crosses) and of Abell 168 cluster and field UDGs from \citet{roman16} (blue  crosses). Diagonal  lines represent constant $\mu_e$ lines. Simulated UDGs with effective radii 1$<$$r_e$/kpc$<$2 are shown as black points, with extreme cases, $r_e$$\geqslant$2 kpc, further circled. The crossed UDG is analysed in \Sec{sec:formation} and \Fig{fig:UDG_evolution}.}
\label{fig:UDG_prop} 
\end{figure}
\vspace{-.3in}\section{Results} \label{sec:results}

Simulated galaxies are defined as UDGs if they satisfy the following criteria: i) their 2D effective radius, $r_e$, is larger than 1 kpc, ii) their absolute magnitude in R band is  $-16.5$$\lesssim$$M_{\rm R}$$\lesssim$$-12$, corresponding to a stellar mass of $10^{7}$$\lesssim$\mstar/\msun$\lesssim$$10^{8.5}$, iii) their effective surface brightness is low, with $\mu_e$$>$23.5 mag/arcsec$^2$. It is worth noticing that several of the NIHAO galaxies fall in the UDGs category. To facilitate comparisons with observational results, we work in the AB system and used UBVR$_{\rm c}$I$_{\rm c}$ Johnson-Cousins filters. Galaxies are face-on (aligned via angular momentum of the stars) when computing $r_e$ and the effective surface brightness is defined as $L/(2\pi r_e^2)$ before converting it in units of mag/arcsec$^2$.

A sample of 21 NIHAO simulations meet these requirements, shown in \Fig{fig:UDG_prop} as black points, with the largest-sized UDGs ($r_e$$\geqslant$2 kpc) further circled.  One of the most extreme UDGs, with  the largest  $r_e$ amongst the lowest surface brightness objects, is  marked with a cross in \Fig{fig:UDG_prop} and it will be  analysed  in \Sec{sec:formation}. In \Fig{fig:UDG_prop} we compare simulations  with observed UDGs. From top to bottom the color B-R, the  2D effective radius and the effective surface brightness are shown against the absolute magnitude $M_{\rm R}$.  SDSS \textit{gri} colors and magnitudes used in \citet{roman16} were converted to BR Subaru-Suprime-Cam ones used in \citet{yagi16} by adopting the color conversions derived in the appendix of \citet{yoshida16}.
When comparing $r_e$ and $\mu_e$ with the \citet{roman16} data we used their r-band results, as best approximation of our R-band ones. 
\begin{table}
\begin{center}
\begin{tabular}{lcc}
\hline
\hline

$X$  & $\overline{X}\pm\sigma$ & X$_{min,max}$ \\
\hline
log(\mstar[\msun])   & 7.66$\pm$0.42 & 6.83 , 8.40    \\                    
log(\mhalo[\msun])   & 10.53$\pm$0.18 & 10.22 , 10.85   \\                    
log(\mHI[\msun])   & 8.37$\pm$0.59 & 7.22 , 9.24   \\                    
$r_e$[kpc]& 1.87$\pm$0.53 & 1.07 , 3.06\\        
$\mu_e$[mag/arcsec$^2$]& 25.23$\pm$0.94 & 23.69 , 26.84 \\        
$M_R$& -14.61$\pm$1.07 & -16.25 , -12.57  \\        
B-R& 0.77$\pm$0.12 & 0.54 , 0.97   \\   
n$_{\rm Sersic}$& 0.83$\pm$0.27 &  0.31 , 1.46 \\   
$\gamma(1$-$2\%R_{vir})$ & -0.37$\pm$0.18  &  -0.78 , -0.01\\
log($\lambda$) & -1.48$\pm$0.25   &  -2.04 , -1.17\\
c$_{\rm DM}$  & 10.67$\pm$3.05  & 5.89,18.85\\
\hline
\hline
\end{tabular}
\label{tab:UDG_sim}
\caption{Average properties of the simulated UDG sample. Concentrations and spin parameters have been computed in the original DM-only run.}
\end{center}
\end{table}

Simulated UDGs overlap with observational data in color, effective radii, surface brightness and magnitude. 
While most of the Coma cluster UDGs follow the red sequence with B-R=0.8-1.2,  the Abell 168 region, including both cluster and field UDGs, span a wider range in colors, with the bluest objects having  B-R$\sim$0.4.
This  broad range in colors is also observed in the simulated UDGs, some of which have recent star formation (see \Sec{sec:gas content}), an indication that not all UDGs are  evolving passively and suggesting that the isolated counterparts of cluster UDGs may not be quenched.

 \subsection{UDGs: global properties of the simulated sample}\label{sec:global}

In Table 1 we summarize the properties  of  simulated UDGs: from top to bottom we  specify  stellar mass, halo mass, HI gas mass (\mHI), 2D effective radius, effective surface brightness, R-band absolute magnitude, B-R color, S\'ersic index, DM  halo inner slope, spin parameter and concentration. Specifically, the S\'ersic index n$_{\rm Sersic}$ is computed by fitting the 2D surface brightness profile in R-band out to  2$\times$r$_e$ with a S\'ersic profile \citep{sersic68}, the inner slope $\gamma$ of the DM halo  is  found by fitting its density profile with a power law between 1 and 2$\%$ of the virial radius, in a region where all our galaxies are well resolved, the dimensionless spin parameter $\lambda$ follows the \citet{bullock01} definition and the concentration $c$ is computed from the original DM-only simulation.

All  the currently observed structural properties  of UDGs (\mstar, n$_{\rm Sersic}$, color, $M_{\rm R}$, $r_e$ and $\mu_e$) are in excellent agreement with the ones of the simulated  sample.
The mean value of the spin parameter is  close to the peak of the distribution of spin parameters for DM haloes (log($\lambda\sim$-1.45), \citealt{bullock01}), indicating that  our simulated UDGs do not live in particularly high-spin objects as suggested by \citet{amorisco16}. 
The range of DM inner slopes, -0.78$<$$\gamma$$<$-0.01, shows that UDGs  live  in expanded DM  haloes, whose logarithmic inner slope is shallower than the universal NFW value of $\gamma$=-1. We will see in \Sec{sec:formation} how this is closely linked to the formation of UDGs. 

Interestingly, the simulated UDGs  have a non-negligible amount of HI gas, whose fraction at z=0 is computed including self-shielding and ionization from star forming regions  as  in \citet{gutcke16}. While most  recent observations  focused on UDGs in  clusters,  finding the not surprising result that those objects are gas-poor, there is no current evidence that isolated UDGs should be gas-poor: indeed, the only work that focused on isolated UDGs  could only place an upper limit of \mHI$<$10$^{8.8}$\msun\ on the expected amount of HI gas in DGSAT I \citep{delgado16}, suggesting that values of 7$\lesssim$log$_{\rm 10}$(\mHI/\msun)$\lesssim$9, as predicted by our simulations, are fully  within observational constraints.
Note that the \mHI\ amount that UDGs should have once falling into a cluster can not be inferred by using our isolated simulations.

 \begin{figure}
\hspace{.5cm}\includegraphics[width=3.in,height=2.in]{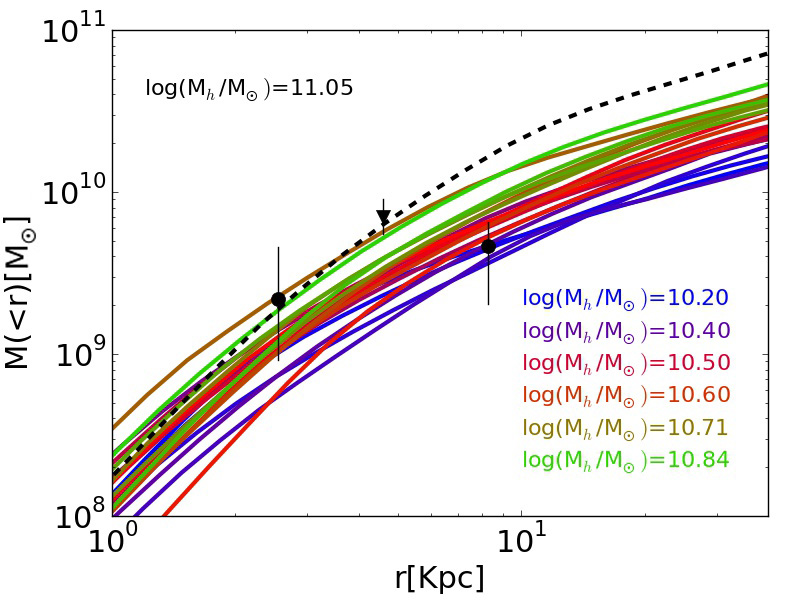}
  \caption{Mass profiles of simulated UDGs, color coded by  DM  halo mass,  including contributions from DM, gas and stars. Overplotted observational results from \citet{beasley16}, VCC 1287 UDG as circles points, and from \citet{vandokkum16}, DF 44 UDG as triangle.}
\label{fig:UDG_mass_prof} 
\end{figure}

 \begin{figure*}\hspace{-1.32cm}\includegraphics[width=2.1in,height=1.6in]{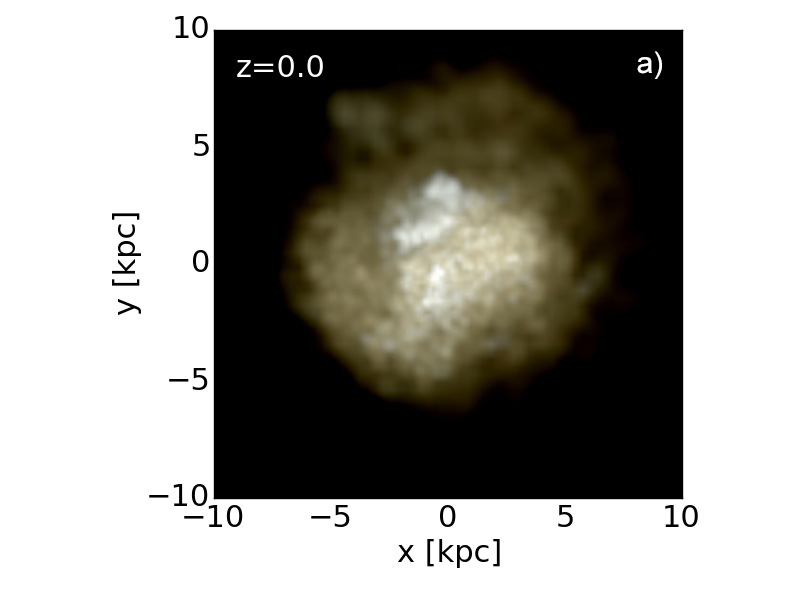}\hspace{-.4cm}\includegraphics[width=1.95in,height=1.75in]{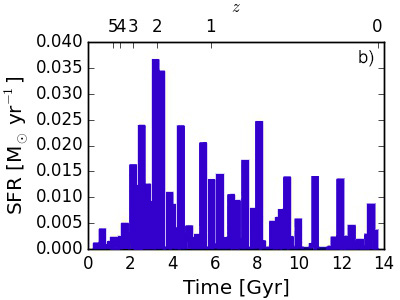}\includegraphics[width=1.8in,height=1.6in]{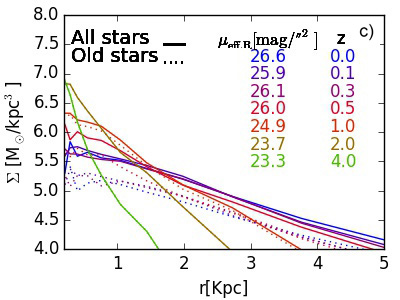}\includegraphics[width=1.8in,height=1.6in]{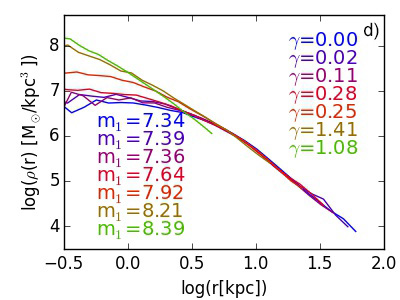}
  \caption{Formation of the UDG marked with a cross in Fig.1. From left to right: a) face-on multi-band image of stars at z=0, b)  SFH, c) 3D spherically averaged stellar density profile color coded by $z$  and d) evolution of DM density profile. Panel c): the contribution of all stars and old stars formed within the first 5 Gyrs of the galaxy's life is indicated as solid and dashed line, respectively; the $\mu_e$  evolution is shown as well. Panel d): the DM density inner slope measured within 1 and 2$\%$ of the galaxy virial radius and the total enclosed mass at 1 kpc (m$_{\rm1}$=log$\rm_{10}(M_{<1kpc}/M_{\odot})$) are indicated as a function of $z$.}
\label{fig:UDG_evolution} 
\end{figure*}
Remarkably, the simulated UDGs have average halo mass \mhalo=10$^{10.53}$\msun\, and a maximum and minimum \mhalo\ of 10$^{10.22}$ and 10$^{10.85}$\msun, respectively: they are therefore legitimate dwarfs rather than failed L$\star$ objects.  Such dwarf halo mass range is also advocated in  \citet{amorisco16}, although their model requires high halo spin, unlike what we see in our simulated UDGs.
    
The halo masses of  the simulated galaxies are in  agreement with the inferred  mass of  UDGs  in the Virgo  \citep{beasley16}  and Coma clusters \citep{Peng16,beasley16b}.
We plot  in \Fig{fig:UDG_mass_prof} their total mass profiles  together with the two available measurements of the  mass of VCC 1287 (circles),  obtaining a halo mass between 10.20$<$log$\rm_{10}$(\mhalo/\msun)$<$10.80, in agreement with estimates from  \citet{beasley16}. 
We  also show DF 44 UDG (triangle) as in \citet{vandokkum16}, who inferred  a $\sim$8$\times$10$^{11}$\msun\   halo mass for this galaxy: we show here that  a \mhalo=10$^{11.05}$\msun\  matches this observation, by plotting as dashed line the mass profile of a simulated UDG with a similar \mstar\ as DF 44 and  effective radius of 4.4 kpc (not shown in Fig.1 due to magnitude cut), corroborating the finding that even the brightest UDGs are not Milky-Way mass objects.      

Finally, the concentrations of simulated UDGs are typical  of galaxies of their halo masses, excluding the possibility that UDGs form in haloes with $c$ systematically  higher or lower  than average.

\subsection{UDGs: formation scenario}\label{sec:formation}    
\vspace{-.1 cm}We analyse the properties of the  UDG marked with a cross in \Fig{fig:UDG_prop} to show how such objects can form. In \Fig{fig:UDG_evolution} we show, a) a visualization of the stellar distribution at z=0 using a 3-color image based on IVU bands, b) the star formation history (SFH) of the galaxy, c) the evolution of the 3D stellar density as a function of redshift for all stars (solid lines) and old stars (t$_{\rm form}$$<$5 Gyrs, dashed lines) and d) the evolution of DM density, logarithmic inner slope $\gamma$ and total amount of mass m$_{\rm1}$=log$\rm_{10}(M_{<1kpc}/M_{\odot})$ contained within 1 kpc of the galaxy  as a function of redshift.
The bursty SFH of  the UDG can be appreciated in panel b), including peaks in the last Gyr of its history, indicating  recent episodes of star formation: this feature is also reflected in the blue  off-centered over-density visible in the colour image of panel a), similar to DGSAT I UDG observed in \citet{delgado16}.\\
\indent The formation scenario of UDGs is illustrated in panels c) and d): as the DM halo expands and forms a central core due to episodic and powerful gas outflows driven by star formation, the stellar distribution  expands as well. In panel c) the effective surface brightness is shown as a function of redshift: as $r_e$ increases, $\mu_e$ decreases bringing the dwarf onto the UDG regime. This  galaxy became an ultra diffuse one by $z$=1.5. To confirm that $r_e$ increases due to expansion of the stellar distribution rather than by new episodes of star formation in the outer regions, we separate the contribution of all stars and old stars: we observe  that even the oldest stellar population expands as a response to the core creation mechanism. A spatially extended stellar distribution, typical of UDGs, can therefore arise from internal feedback processes, which also give rise to a spherical rather than disky galaxy. An extensive study of UDGs morphologies will be presented in forthcoming work.

\subsection{UDGs: gas content and star formation history}\label{sec:gas content}      
\begin{figure}
\hspace{-.4cm}\includegraphics[width=3.6in,height=4in]{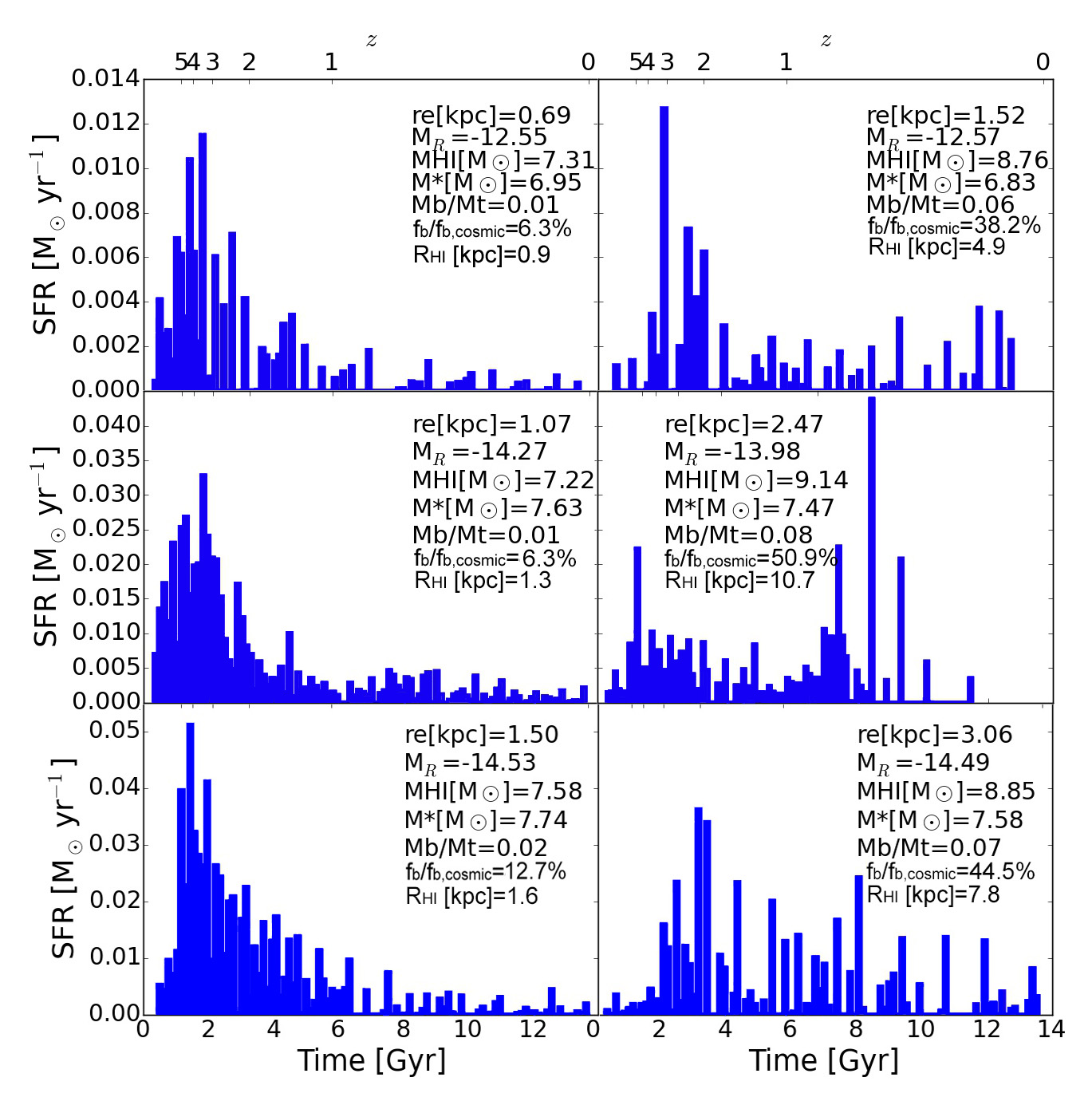}\caption{SFHs of galaxies with the largest (right column) and smallest (left column) effective radius in their mass bin. From top to bottom, each row shows galaxies with similar halo mass and magnitude, log$_{10}$(\mhalo/\msun)$\sim$10.20,10.45,10.50 and  $M_{\rm R}$$\sim$-12.5,-14.0,-14.5. In each panel  $r_e$, $M_{\rm R}$, \mHI, \mstar, f$_b$ and HI radius are indicated.}
\label{fig:SFHs} 
\end{figure}

\vspace{-.0cm}We  investigate  what makes UDGs differ from more compact galaxies in the same stellar mass range.
In \Fig{fig:SFHs} we show the SFH of six galaxies whose effective radii are the largest (right column) and smallest (left column) in their respective mass bin. \Fig{fig:SFHs}  therefore includes both UDGs as well as more compact, regular dwarfs.
From top to bottom, we pair galaxies with  similar halo  and  stellar masses, quoting in each panel the $r_e$, $M_R$, \mstar, \mHI, extension of HI gas (R$\rm_{HI}$, as the radius at which the HI surface density reaches  1 \msun/pc$^2$) and baryon fraction relative to the cosmic one,  f$_{b,c}$.\\
\indent The difference in properties between the most extreme UDGs (right panels) and the less extreme, more compact dwarfs (left panels) are striking. Galaxies with large $r_e$ also have a larger \mHI, baryon fraction and R$\rm_{HI}$, and more prolonged and persistently bursty SFH, including a larger fraction of young stellar population, compared to galaxies with a smaller $r_e$. When most  of star formation happens in the first $\sim$3-4 Gyrs, feedback  can eject significant amounts of gas from relatively shallow potential wells at early times \citep[e.g.][]{dekel86}, resulting in low baryon fractions by $z$=0. 
Since gas is expelled at early stages, there is less gas for ongoing star formation and crucially there is less gas to be expelled from the inner regions when star formation occurs, being this the key aspect of the mechanism for core creation: the lower is the gas fraction at a given epoch, the less efficient is such mechanism. We verified that the dwarfs with lower $r_e$ have retained at most $10\%$ of the initial gas mass between $z$=4 and 1, while the largest $r_e$ UDGs have kept up to $50\%$ of the initial gas within the same period.
The less expanded galaxies have a very low baryon fraction, f$_b$/f$_{b,c}$$\sim$6-13$\%$ by $z$=0, and retain up to two order of magnitudes less HI gas than similar \mstar\ galaxies with larger $r_e$; their DM inner slope is less shallow and correspondingly the distribution of HI is more compact, with R$\rm_{HI}$$\sim$0.9-1.6 kpc.\\
\indent Oppositely, galaxies with star bursts occurring after the rapid halo growth phase has finished are the ones that can keep their gas, which can not escape the deeper potential well: they have enough gas available at all time to drive DM cores and a spatially extended stellar distribution,  retaining about $50\%$f$_{b,c}$ and up to 10$^9$\msun\ in HI gas by $z$=0. A similar dependence of core sizes with SFH has been found in \citet{onorbe15} and \citet{read16} for lower mass objects than the ones studied here, with \mhalo$\sim$10$^{7-10}$\msun.\\
We conclude by summarizing the differences between regular dwarfs (such as the top-left galaxy in \Fig{fig:SFHs}) and UDGs within a similar \mstar\ range: non-UDGs have smaller effective radius, less gas mass and baryon fraction, steeper DM inner slope and a higher S\'ersic index than UDGs, while their halo mass, color, magnitude, spin parameter and concentration are indistinguishable from the ones of UDGs. This further validates the finding that the   availability of gas is crucial to the  formation mechanism of UDGs.

\vspace{-.3cm}\section{Conclusions} \label{sec:discussion}
We showed that cosmological simulations of isolated galaxies from the NIHAO project,  which include feedback from SNe and massive stars, reproduce a population of Ultra-Diffuse Galaxies (UDGs) with stellar mass, magnitude, colour, S\'ersic index, effective radius and surface brightness in  agreement with observations. Internal processes, rather than environmental ones, are at the base of the formation scenario of  UDGs. Feedback driven gas outflows give rise to a spatially extended stellar component, while simultaneously expanding the dark matter halo, leading to the emergence of low surface brightness dwarf galaxies, or UDGs, with \mstar$\sim$10$^{7-9}$\msun\ and $r_e$$\sim$1-3 kpc. A key aspect is the availability of gas that is driven away from central regions during bursts of star formation, causing rapid oscillations of the potential: galaxies that expel most of their gas at early times are less efficient at causing expansion than galaxies of similar \mstar\ that retain more gas for later times. Our findings imply that UDGs:
\begin{itemize}
\item  are  dwarf galaxies, with $10^{10}$$\lesssim$\mhalo/\msun$\lesssim$$10^{11}$, in agreement with mass estimates from \citet{beasley16}, \citet{beasley16b} and \citet{Peng16}, and with our revised mass estimate of the \citet{vandokkum16} DF 44 UDG: they  naturally extend the population of  LSB galaxies to the dwarf  regime;
\item  are not exclusively associated to a cluster environment: they are  expected to be found in the field as well; 
\item  are not all red and quenched: we found simulated UDGs with B-R$<$0.7 and recent off-centered star formation, as well as UDGs with B-R$>$0.7 that only stopped forming stars in the past 2 Gyrs; 
\item have typical distributions of halo spin and concentration, an average S\'ersic index of less than one and dark-matter cores;
\item  if isolated, have significant HI gas mass, \mHI$\sim$10$^{7-9}$\msun: at a similar stellar mass, the larger the effective radius, the higher is the baryon fraction retained within the virial radius and the larger is the amount and extent of HI gas, with R$\rm_{HI}$ up to $\sim$10.7 kpc; moreover, the larger  $r_e$, the higher is the fraction of young stellar population expected in isolated UDGs.
\end{itemize}

\noindent The first three points of the list are in common with the model of \citet{amorisco16} although, unlike our scenario, those authors  appealed to high spin haloes in order to form UDGs.

Given their predicted  \mHI,  some  UDGs should be seen, or have already been seen, by the  HI Alfalfa survey \citep{alfalfa1,haynes11}.
Indeed, some \textit{dark galaxies} in Alfalfa, those without  a clear optical counterpart, have only recently been identified with LSB objects  \citep{cannon15} and  have HI mass 7.4$<$log(\mHI/\msun)$<$9.5.
Deep optical imaging \citep{janowiecki15} shows that the system HI1232+20 has magnitude, HI mass, surface brightness, colors and remarkably a large R$\rm_{HI}$ (5.1 to 11.2 kpc)  in agreement with our most expanded UDGs. 
The fraction of \textit{dark galaxies} over the total number of detected HI sources, within a  completeness radius of 20 Mpc and having the expected \mHI\ mass of UDGs, is as high as $8.5\%$ in the Alfalfa $\alpha$.70 catalogue: some of them may be UDGs, and future dedicated  observations  can help verify our claim.
Finally, in the SPARC sample of local HI-rich galaxies \citep{lelli16}, 70$\%$ of the galaxies with 10$^{7}$$<$\mstar/\msun$<$10$^{8.5}$  have $r_e$$>$1 kpc whilst 20$\%$ have  $r_e$$>$2 kpc, giving surface brightnesses in the realm of UDGs, and showing that such sizes are not  rare amongst dwarfs after all.\\

We proposed a scenario in which UDGs form naturally by outflow episodes in  haloes of \mhalo$\sim$$10^{10-11}$\msun, in the same mass range where  feedback induced expansion is expected to be most efficient at creating a cored distribution of DM and stars \citep[e.g.][]{DiCintio2014a,tollet15,dutton16}: the existence of a preferential halo mass for UDGs therefore fits  within our theoretical understanding of how stellar feedback  impacts the DM and stellar distribution in galaxies.
This picture  agrees with  models in which the effect of  repeated  outflows accumulates during cosmic time \citep[e.g.][]{read05,pontzen12,dutton16}: we demonstrate here for the first time that it is the gas availability  that drives both  the  SFH and the amount of expansion of  DM and stars, with
 higher gas fractions  more efficiently expanding both the stellar and DM component, leading to  the emergence of low surface brightness, gas-rich, HI extended, UDGs.  

Larger diffuse galaxies, with $r_e$ up to 7-8 kpc, exist as well in the NIHAO simulations, with \mstar\ higher than the presently observed UDGs: they can be classified as regular LSBs and will be the subject of a future paper (Di Cintio in prep).

\vspace{-.5cm}\section*{Acknowledgements}
The authors kindly thank M. Yagi, J. Koda, I. Trujillo, J. Roman, M. Beasley, E. Papastergis, M. Haynes, F. Lelli and N. Amorisco for sharing their data and for fruitful discussions. Computational resources  were provided by the High Performance Computing at NYUAD, the  {\sc theo} cluster at MPIA and  the {\sc hydra} clusters  at Rechenzentrum in Garching. ADC thanks the Carlsberg foundation. CBB thanks MINECO/FEDER  grant AYA2015-63810-P. AD is supported by grants ISF 124/12, BSF 2014-273.
\vspace{-.4cm}
\bibliographystyle{mn2e}
\bibliography{archive}


\label{lastpage}

\end{document}